\documentclass[useAMS,referee,usenatbib]{biom}
\usepackage{url}
\usepackage{amsmath}

\def\bSig\mathbf{\Sigma}

\usepackage[figuresright]{rotating}

\title[Unified Inference Method for FROC Data]{A Unified Inference Method for FROC-type Curves and Related Summary Indices}

\author
{Jiarui Sun$^{1}$, 
Kaiyuan Liu$^{1}$ and 
Xiao-Hua Zhou $^{2,*}$\email{azhou@math.pku.edu.cn}, \\
$^{1}$School of Mathematical Sciences, Peking University, Beijing, China \\
$^{2}$Beijing International Center for Mathematical Research and Department of Biostatistics,\\ Peking University, Beijing, China}

\begin{document}


\date{{\it Received XX} XXXX. {\it Revised XX} XXXX.  {\it
Accepted XX} XXXX.}



\pagerange{\pageref{firstpage}--\pageref{lastpage}} 
\volume{XX}
\pubyear{XXXX}
\artmonth{XXXX}


\doi{XXXX}


\label{firstpage}


\begin{abstract}
Free-response observer performance studies are of great importance for accuracy evaluation and comparison in tasks related to the detection and localization of multiple targets or signals. The free-response receiver operating characteristic (FROC) curve and many similar curves based on the free-response observer performance assessment data are important tools to display the accuracy of detection under different thresholds. The true positive rate at a fixed false positive rate and summary indices such as the area under the FROC curve are also commonly used as the figures of merit in the statistical evaluation of these studies. Motivated by a free-response observer performance assessment research of a Software as a Medical Device (SaMD), we propose a unified method based on the initial-detection-and-candidate model to simultaneously estimate a smooth curve and derive confidence intervals for summary indices and the true positive rate at a fixed false positive rate. A maximum likelihood estimator is proposed and its asymptotic normality property is derived. Confidence intervals are constructed based on the asymptotic normality of our maximum likelihood estimator. Simulation studies are conducted to evaluate the finite sample performance of the proposed method. We apply the proposed method to evaluate the diagnostic performance of the SaMD for detecting pulmonary lesions.

\end{abstract}

%



\maketitle


%

\section{Introduction}
\label{s:intro}

Free-response observer performance assessment is widely used to evaluate or compare detection and localization accuracy for multiple targets or signals. It plays an important role in signal detection \citep{miller1969froc} and diagnostic radiology \citep{chakraborty2017observer}. For example, in diagnostic radiology, it is important to not only identify diseased patients but also to find the location of disease \citep{winawer2007colorectal,black2000anatomic}. Especially in the field of Software as a Medical Device (SaMD), location information provided by the free-response observer assessment is crucial to the verification of the accuracy of the medical device. Motivated by the free-response observer performance study of a SaMD introduced in Section~\ref{s:application}, we propose a new statistical inference method for the detection and localization accuracy evaluation based on the free-response observer performance assessment data. We describe our method under the application scenario of diagnostic radiology, but our method can also be used for general free-response observer performance studies.

In general, there are three types of accuracy evaluation indices based on the free-response observer performance assessment data. The first type of indices is the characteristic curve that is similar to the receiver operating characteristic (ROC) curve (see \citet{zhou2009statistical} for the definition of a ROC curve), which displays the performance of the evaluated diagnostic test under different thresholds. Examples are the free-response receiver operating characteristic (FROC) curve  \citep{bunch1977free}, the alternative free-response receiver operating characteristic (AFROC) curve \citep{chakraborty1990free}, and the exponential transformation free-response receiver operating characteristic (EFROC) curve \citep{popescu2011nonparametric}. We call them FROC-type curves for short in this article. The FROC-type curves provide direct visual representations of accuracy data and do not require the selection of a particular decision threshold. The second type of indices is the summary index which characterizes the overall performance of the diagnostic method, such as the areas under the just mentioned curves, the partial area under the FROC curve \citep{samuelson2006comparing}, and the summary index proposed in \citet{bandos2009area}. A summary index summarizes the accuracy of a test by a single number and is useful for the comparison of different diagnostic tools. The third type of indices measures the true positive rate at a fixed false positive rate, which is similar to the sensitivity at a fixed specificity in ROC studies. In clinical practice, the false positive rate has to be constrained below a certain value due to the consequences of misdiagnoses. As a result, we usually control the false positive rate at a fixed value and evaluate the corresponding true positive rate. For example, the ordinate at a given abscissa on the FROC curve is the index considered in \citet{nagel1998analysis}. The true positive rate at a fixed false positive rate is easy to interpret and is preferred when evaluating a test for a particular application \citep{zhou2009statistical}. These three types of accuracy indices for the FROC data are equally important and are used for different study purposes. We focus on the estimation and inference of these three types of indices in this article.

There are some existing methods for the estimation and inference of the accuracy evaluation indices mentioned above, but each of them has its own limitation. In previous research, the empirical method is commonly used for the estimation of FROC-type curves and summary indices. With this method, empirical FROC-type curves are constructed by connecting some discrete points formed by varying thresholds. The corresponding AUCs are also estimated by the area under the empirical curves \citep{chakraborty2017observer,chakraborty2004observer,chakraborty2008validation}. 
Such empirical curves are not smooth while the true FROC-type curves are smooth and usually convex. As a result, the FROC-type curves can be underestimated by the empirical method, thus leading to the underestimation of the diagnostic accuracy. The confidence intervals for summary indices such as AUCs are often constructed by the bootstrap method \citep{samuelson2006comparing,bandos2009area}. As far as we know, there are no asymptotic normality results of empirical estimates for AUCs of FROC-type curves, and hence the confidence intervals derived through the bootstrap method may lack theoretical justification. Furthermore, no inference method has been proposed for the true positive rate at a fixed false positive rate. Although some parametric methods aiming at estimating a smooth curve are available  \citep{edwards2002maximum,chakraborty1989maximum,chakraborty2017observer,chakraborty2006search}, these methods mainly focus on the point estimation without providing the calculation method of variance so inference problems such as the construction of confidence intervals and hypothesis testing can not be done. In summary, a new statistical method is needed for valid estimation and inference of the three types of accuracy evaluation indices.


In this article, we propose a unified method to simultaneously estimate a smooth FROC-type curve, derive the confidence intervals for summary indices such as AUC, and derive the confidence interval for the true positive rate at a fixed false positive rate. By `unified' we mean that our method can deal with the estimation of a smooth curve and the inference of the indices at the same time, while existing methods can only deal with one of them. Our method has several advantages. First, it can simultaneously estimate a smooth curve and derive the confidence intervals for summary indices like AUC. The confidence intervals of our method are based on the asymptotic normality results and have a theoretical coverage rate guarantee compared to the existing methods. Second, our method can derive the confidence interval for the true positive rate at a fixed false positive rate for FROC data, which can't be done by any existing FROC method as far as we know. Finally, our method can be used for the inference of multiple summary indices. Exsiting methods mostly deal with the inference problem of a single index without providing the confidence sets for multiple indices. A single index is not always sufficient for the accuracy evaluation. Different indices may contain information on different aspects of the evaluated diagnostic method. We illustrate the necessity for using multiple indices with a real example in Section \ref{s:application} in more detail.

The remainder of the article is organized as follows. Section~\ref{s:notation} describes the notations used in this article and proposes the model assumptions. Section~\ref{s:result} provides the main results of our methods, including the asymptotic property of our MLE estimator and the construction of the confidence intervals. Section~\ref{s:simulation} is a simulation study and Section~\ref{s:application} is our application on a clinical trial of a Software as a Medical Device (SaMD). Finally, a discussion is given in Section \ref{s:discuss}.

\section{Notations and Model Assumptions}
\label{s:notation}

In this section, we introduce the notations and the model assumptions. For narrative convenience, we assume that the scenario in our article is a diagnostic radiology study, where readers are required to detect and locate lesions on the images of patients.

\subsection{Notations}

Consider an FROC study with a total number of $N=K_1+K_2$ subjects. The gold standard records the locations of lesions on each subject. A subject may have one lesion, multiple lesions, or no lesion at all. If a subject has at least one lesion, then it is classified as a positive subject. If a subject has no lesion, it is classified as a negative subject. Hence, N subjects can be divided into $K_1$ positive subjects and $K_2$ negative subjects. For positive subjects, denote $t_i$ as the number of lesions on the ith positive subject, where $i=1,2,\dots,K_1$ and $t_i \geq 1$. As a consequence, there are totally $T=\sum_{i=1}^{K_1}t_i$ lesions.

In an FROC study, each reader is asked to record all of the suspicious locations and assign the corresponding confidence score to each suspicious location. The reader may record multiple suspicious locations on one subject and may also record no location. The confidence score reflects the confidence for classifying a recorded location as a lesion. The higher the confidence score, the stronger the confidence. The suspicious locations recorded by the reader are then compared to those of the gold standard. If a suspicious location recorded by the reader matches with one of the locations recorded by the gold standard, this suspicious location is a true positive (TP). If a suspicious location recorded by the reader can not match with any of the locations recorded by the gold standard, this suspicious location is a false positive (FP).

Here we introduce the notations used in this article. Denote $L_{is}$ as the binary variable indicating whether the sth lesion on the ith positive subject is recorded by the reader, where $i=1,\dots,K_1$, and $s=1,2,\dots,t_i$. Denote $n_i$ as the number of false positive locations on the ith positive subject and $m_j$ as the number of false positive locations on the jth negative subject, where $j=1,\dots,K_2$. The confidence score of a true positive location is denoted as $Y_{is}$ for i and s such that $L_{is}=1$. If $L_{is}=0$, corresponding $Y_{is}$ does not exist, because this lesion is not recorded by the reader. We denote the confidence score of the false positive locations on the ith positive subject as $X^*_{iq}$, where $q = 0,1,\dots,n_i$. If $n_i=0$, there are no false positive locations on the ith positive subject. Denote $X_{jr}$ as the confidence scores of the locations on the jth negative subject, where $r=0,1,\dots,m_j$. If $m_j=0$, there are no false positive locations on the jth negative subject. The FROC data can be summarized as follows:
\[
	\left\{t_i,\left\{L_{is}\right\}_{s=1}^{t_i},
	\left\{Y_{is}\right\}_{s=1}^{t_i},n_i,
	\left\{X^*_{iq}\right\}_{q=0}^{n_i}\right\}_{i=1}^{K_1},
\]
\[
	\left\{m_j,
	\left\{X_{jr}\right\}_{r=0}^{m_j},
	\right\}_{j=1}^{K_2}.
\]
Note that if $L_{is}=0$, $Y_{is}$ does not exist. If $n_i=0$, $\left\{X^*_{iq}\right\}_{q=0}^{n_i}$ is an empty set. If $m_j=0$, $\left\{X_{jr}\right\}_{r=0}^{m_j}$ is an empty set.

\begin{figure}
	\centerline{\includegraphics[width=2.25in]{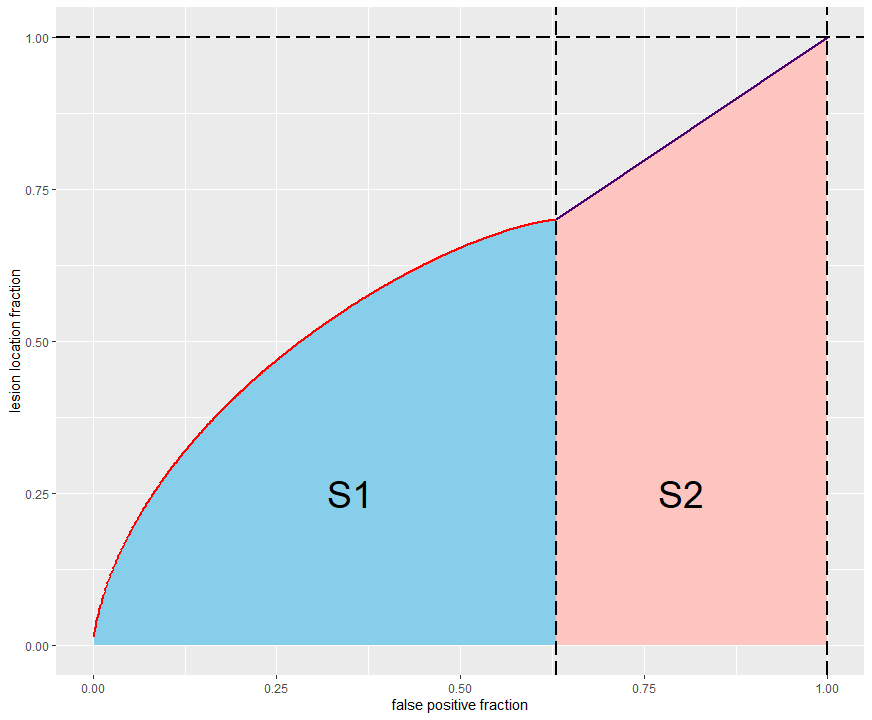}}
	\caption{An example of an AFROC curve}
	\label{f:AFROCexample}
\end{figure}

Under the just introduced notations, the lesion localization fraction (LLF) and the false positive fraction (FPF) at threshold $\zeta$ can be defined as:
\begin{equation}
	FPF(\zeta)=P(\max \limits_{1\leq r \leq m_j}X_{jr}>\zeta;m_j>0),
	\label{equ:FPF}
\end{equation}	
and
\begin{equation}
	LLF(\zeta)=P(Y_{is}>\zeta;L_{is}=1),
	\label{equ:LLF}
\end{equation}
respectively. The alternative FROC (AFROC) curve is the plot of $LLF(\zeta)$ versus $FPF(\zeta)$ as $\zeta$ varies from $-\infty$ to $+\infty$. It can be equivalently expressed as 
\[AFROC(t)=LLF(FPF^{-1}(t)), \]
where t ranges from 0 to $P(m_j>0)$. An example of an AFROC curve is shown in Figure \ref{f:AFROCexample}. 

The area under the AFROC curve (AUC) consists of two parts, which we denote as $S_1$ and $S_2$ in Figure \ref{f:AFROCexample}. $S_1$ is the area under the curve formulated in Equation~\ref{equ:FPF} and Equation~\ref{equ:LLF}. $S_2$ is the area under the straight line connecting $(P(m_j>0),P(L_{is}=1))$ and $(1,1)$. AUC can be written as follows:
\[
AUC=S_1+S_2=\int_{t=0}^{P(m_j>0)}LLF(FPF^{-1}(t))dt+\frac{(P(L_{is}=1)+1)P(m_j=0)}{2}
\]
\[
	=P(\max \limits_{1\leq r \leq m_j}X_{jr}<Y_{is};m_j>0,L_{is}=1)+\frac{(P(L_{is}=1)+1)P(m_j=0)}{2}.
\]

\subsection{Model Assumptions}

We consider a continuous version of the commonly used initial-detection-and-candidate-analysis model (IDCA) \citep{edwards2002maximum}. The IDCA model divides the diagnostic process of a computer observer into two stages. In the first stage, the observer views the whole image and selects the suspicious locations, which are called initial candidate detections. In the second stage, each suspicious location recorded by the observer in the first stage is given a confidence score. This model describes the procedure of the search work for a computer observer (such as a SaMD), where it first locates a large number of candidate locations and then analyzes each candidate location. 

For this two stage IDCA model, we make the following model assumptions:

$\bm{A}_1$: $L_{is}$ follows a Bernoulli distribution with $P(L_{is}=1)=p$. $n_i$ follows a poisson distribution with parameter $\lambda_2$. $m_j$ follows a poisson distribution with parameter $\lambda$. That's to say, $P(n_i=k)=\frac{\lambda_2^k}{k!}e^{-\lambda_2}$ and $P(m_j=k)=\frac{\lambda^k}{k!}e^{-\lambda}$.

$\bm{A}_2$:  $L_{is}$, $n_i$, $m_j$ are mutually independent.

$\bm{A}_3$: Conditioning on $\{L_{is}\}_{i,s}$, $Y_{is}$ are independent (if exist) and follow a distribution with cumulative distribution function (cdf) $G_{\bm{\theta}_1}$. $\bm{\theta}_1$ is an unknown constant (or vector) which is not associated with $L_{is}$, $n_i$ or $m_j$.

$\bm{A}_4$: Conditioning on $\{m_j\}_{j}$, $X_{jr}$ are independent (if exist) and follow a distribution with cdf $F_{\bm{\theta_2}}$. Conditioning on $\{n_i\}_{i}$, $X^*_{iq}$ are independent (if exist) and follows a distribution with cdf $F_{\bm{\theta_3}}$. $\bm{\theta}_2$ and $\bm{\theta}_3$ are unknown constants (or vectors), which are not associated with $L_{is}$, $n_i$ or $m_j$. $F_{\bm{\theta}_2}$ and $F_{\bm{\theta}_3}$ need not to be the same distribution in our method. 

$\bm{A}_5$: $\{Y_{is}\}_{i,s}$, $\{X^*_{iq}\}_{i,q}$ and $\{X_{jr}\}_{j,r}$ are independent.

$\bm{A}_6$: $G_{\bm{\theta}_1}$, $F_{\bm{\theta_2}}$, and $F_{\bm{\theta_3}}$ are continuous distributions. The Fisher information matrices are positive definite for $G_{\bm{\theta}_1}$, $F_{\bm{\theta_2}}$, and $F_{\bm{\theta_3}}$. $G_{\bm{\theta}_1}$, $F_{\bm{\theta_2}}$, and $F_{\bm{\theta_3}}$ satisfy
\[
\frac{\partial}{\partial \bm{\theta}}\int \phi_{\bm{\theta}}(x)dx=\int \frac{\partial}{\partial \bm{\theta}} \phi_{\bm{\theta}}(x)dx
\]
for $\phi_{\bm{\theta}}=f_{\bm{\theta}}(x)$ and $=\partial f_{\bm{\theta}}(x)/\partial \bm{\theta}$, where $f$ is the density function for $G_{\bm{\theta}_1}$, $F_{\bm{\theta_2}}$, or $F_{\bm{\theta_3}}$ and $\bm{\theta}$ is the corresponding distribution parameter.

Denote $\mathcal{Y}_{is}=(L_{is},Y_{is})$ if $L_{is}=1$ and $\mathcal{Y}_{is}=(L_{is})$ if $L_{is}=0$. Denote $\mathcal{X}_{j}=(m_j,X_{j1},\dots,X_{jm_j})$ if $m_j>0$ and $\mathcal{X}_{j}=(m_j)$ if $m_j=0$. 
Denote $\bm{\psi}=(p,\bm{\theta}_1)$. Assume that $\bm{\theta}_1 \in \Theta_1$ and $\Theta_1$ is an open set in $\mathcal{R}^{s_1}$. Denote $\bm{\gamma}=(\lambda,\bm{\theta}_2)$. Assume that $\bm{\theta}_2 \in \Theta_2$ and $\Theta_2$ is an open set in $\mathcal{R}^{s_2}$. As a consequence, $\bm{\gamma} \in \Gamma=\mathcal{R}^+ \times\Theta_2 $ and $\bm{\psi} \in \Psi=(0,1) \times\Theta_1 $. Denote $l^{(1)}_{\bm{\psi}}(\mathcal{Y}_{is})=p^{L_{is}}(1-p)^{1-L_{is}}(g_{\bm{\theta}_1}(Y_{is}))^{L_{is}}$, which is the likelihood function for $\mathcal{Y}_{is}$. Denote $l_{\bm{\gamma}}^{(2)}(\mathcal{X}_j)=\frac{\lambda^{m_j}}{m_j!}e^{-\lambda}\Pi_{r=1}^{m_j}f_{\bm{\theta}_2}(X_{jr})$, which is the likelihood function for $\mathcal{X}_j$. 

$\bm{A}_7:$ For any given $\bm{\gamma} \in \Gamma$, there exist a positive number $c_{\bm{\gamma}}$ and a positive function $h_{\bm{\gamma}}$ such that $E[h_{\bm{\gamma}}(\mathcal{X}_j)]<+\infty$ and 
\[
\sup_{\bm{\xi}:\Vert \bm{\xi}-\bm{\gamma} \Vert <c_{\bm{\gamma}}}\Vert\frac{\partial^2 log l^{(2)}_{\bm{\gamma}}(\mathcal{X})}{\partial\bm{\gamma} \partial \bm{\gamma}^T} \Vert\leq h_{\bm{\gamma}}(\mathcal{X})
\]
for every possible value $\mathcal{X}$ in the range of $\mathcal{X}_1$, where $\Vert A\Vert=\sqrt{\text{tr}(A^TA)}$. For any given $\bm{\psi} \in \Psi$, there exist a positive number $d_{\bm{\psi}}$ and a positive function $m_{\bm{\psi}}$ such that $E[m_{\bm{\psi}}(\mathcal{Y}_{is})]<+\infty$ and 
\[
\sup_{\bm{\xi}:\Vert \bm{\xi}-\bm{\psi} \Vert <d_{\bm{\psi}}}\Vert\frac{\partial^2 log l^{(1)}_{\bm{\psi}}(\mathcal{Y})}{\partial\bm{\psi} \partial \bm{\psi}^T} \Vert\leq m_{\bm{\psi}}(\mathcal{Y})
\]
for every possible value $\mathcal{Y}$ in the range of $\mathcal{Y}_{11}$, where $\Vert A\Vert=\sqrt{\text{tr}(A^TA)}$.

While the original IDCA model was proposed for ordinal confidence scores, we focus on continuous confidence scores. The continuous confidence scores are common for computer observer standalone data. Assumption $\bm{A}_1$ is a natural assumption and has been widely used in search models such as \cite{chakraborty1989maximum} and \cite{edwards2002maximum}. Assumption $\bm{A}_2$ assumes that the number of true positive (TP) locations and the number of false positive (FP) locations on the ith subject are independent of the TP location number and FP location number on the jth subject ($i\neq j$), which is generally true because they are the numbers of locations on different subjects. Assumption $\bm{A}_2$ also assumes that the TP location number and FP location number on the same subject are independent. Assumption $\bm{A}_3$, $\bm{A}_4$ and $\bm{A}_5$ assume that conditioning on the TP location numbers and FP location numbers, both confidence scores of locations on the same subject and confidence scores of locations on different subjects are independent. These independence assumptions are criticised in \citet{bandos2009area} to be difficult-to-justify. However, as \citet{edwards2002maximum} explained that `model assumption regarding the independence of observer detections might be less radical for computer vision methods than for human observers', which is the scene in our application. Computer observers are not human readers. Their diagnostic results are produced by algorithms and the diagnostic result on one location is not likely to affect other locations. In addition, we conduct a simulation study in Section \ref{s:simulation} to assess the effect of the independence assumption on our results. The results show that under mild dependence, our method still has good performance. Assumption $\bm{A_6}$ is easy to check for many commonly used distributions like the exponential family. Assumption $\bm{A}_7$ can be checked by direct calculation. Validation of Assumption $\bm{A}_7$ for some distributions are shown in Supporting Information Web Appendix A. 


\section{Main Results}
\label{s:result}

We first derive the maximum likelihood estimator (MLE) for $(p,\lambda,\bm{\theta}_1,\bm{\theta}_2)$ and its asymptotic property based on the IDCA model. Then we use the MLE estimator for $(p,\lambda,\bm{\theta}_1,\bm{\theta}_2)$ to derive the estimator of accuracy indices and the corresponding asymptotic property. The likelihood function is
\[
	\mathcal{L}(p,\lambda,\bm{\theta}_1,\bm{\theta}_2)=\mathcal{L}_1(p,\bm{\theta}_1)\mathcal{L}_2(\lambda,\bm{\theta}_2),\]
where	
\[	\mathcal{L}_1(p,\bm{\theta}_1)=\Pi_{i=1}^{K_1}\Pi_{s=1}^{t_i}\{p^{L_{is}}(1-p)^{1-L_{is}}(g_{\bm{\theta}_1}(Y_{is}))^{L_{is}}\},\]
and
\[
		\mathcal{L}_2(\lambda,\bm{\theta}_2)=\Pi_{j=1}^{K_2}\{\frac{\lambda^{m_j}}{m_j!}e^{-\lambda}\Pi_{r=1}^{m_j}f_{\mathbf{\bm{\theta}}_2}(X_{jr})\}. 
\]

We define $\Pi_{r=1}^{0}f_{\bm{\theta}_2}(X_{jr})$ to be 1. 

The MLE estimator is derived by maximizing $ \mathcal{L}(p,\lambda,\bm{\theta}_1,\bm{\theta}_2)$. For example, if we assume $\bm{\theta}_i=(\mu_i,\sigma_i)$ and $f_{\bm{\theta}_i}\sim \mathcal{N}(\mu_i,\sigma_i^2)$, then the MLE estimator is $\widehat{p}=\frac{\sum_{i=1}^TL_i}{T}$, $\widehat{\lambda}=\frac{\sum_{j=1}^{K_2}m_j}{K_2}$, $\widehat{\mu}_1=\frac{\sum_{i=1}^{T}\sum_{s=0}^{L_i}Y_{is}}{\sum_{i=1}^{T}L_i}$,\quad $\widehat{\mu}_2=\frac{\sum_{j=1}^{K_2}\sum_{r=0}^{m_j}X_{jr}}{\sum_{j=1}^{K_2}m_j}$, $\widehat{\sigma}_1=\sqrt{\frac{\sum_{i=1}^T\sum_{s=0}^{L_i}(Y_{is}-\widehat{\mu}_1)^2}{\sum_{i=1}^{T}L_i}}$ and $\widehat{\sigma}_2=\sqrt{\frac{\sum_{j=1}^{K_2}\sum_{r=0}^{m_j}(X_{jr}-\widehat{\mu}_2)^2}{\sum_{j=1}^{K_2}m_j}}$. For a more complex distribution, MLE estimator can be obtained through optimization algorithms.

\subsection{Asymptotic Property for the MLE Estimator}

When we derive the asymptotic distribution for the MLE estimator $(\hat{p},\hat{\lambda},\hat{\bm{\theta}}_1,\hat{\bm{\theta}}_2)$, we are faced with a problem of random sample size. In the IDCA model, p and $\lambda$ are related to the distributions of integer-valued random variables $L_{is}$ and $m_j$, while $\sum_{i,s} L_{is}$ and $\sum_j m_j$ are the sample sizes of variables $Y_{is}$ and $X_{jr}$ whose distributions are decided by unknown parameters $(\bm{\theta}_1,\bm{\theta}_2)$. If we are only interested in p, $\lambda$, or if we are only interested in $\bm{\theta}_1$ and $\bm{\theta}_2$, it can be viewed as a fixed sample size problem with sample sizes $\{T,K_2\}$ and $\{\sum_{i,s} L_{is},\sum_j m_j\}$ respectively. However, since the parameter of interest, $AUC=g(p,\lambda,\bm{\theta}_1,\bm{\theta}_2)$, is related to p, $\lambda$, $\bm{\theta}_1$ and $\bm{\theta}_2$, we need to consider the joint distribution of the estimators. Clearly, the variances of the estimator $\widehat{\bm{\theta}}_1$ and $\widehat{\bm{\theta}}_2$ are associated with the sample sizes $\sum_{i,s} L_{is}$ and $\sum_j m_j$, thus related to p and $\lambda$, so there is a problem of `random sample size'. The regular assumption of independently identically distribution for the asymptotic property of MLE estimator breaks down, which brings difficulty for statistical inference. 




Theorem~\ref{asmptoticprop} below gives the asymptotic property for the MLE estimator based on the IDCA model.

\begin{theorem}
	\label{asmptoticprop}
	Under Assumption $\bm{A}_1$ to $\bm{A}_7$, there is a sequence of maximum likelihood estimators of $(p, \lambda, \bm{\theta}_1,\bm{\theta}_2)$, denoted as $(\hat{p}, \hat{\lambda}, \hat{\bm{\theta}}_1,\hat{\bm{\theta}}_2)$, that is consistent and has the asymptotic distribution of

	\[
\sqrt{K_2}\left( 
\begin{array}{c}   
	\widehat{\lambda}-\lambda \\  
	\widehat{p}-p\\
	\widehat{\bm{\theta}}_2-\bm{\theta}_2 \\  
	\widehat{\bm{\theta}}_1-\bm{\theta}_1
\end{array}
\right)\xrightarrow{d}\mathbf{N}\left(\bm{0},\left( 
\begin{array}{cccc}   
	\mathbf{I}_1(\lambda)^{-1}& \bm{0}&\bm{0} &\bm{0} \\  
	\bm{0}&  \frac{1}{c}\mathbf{I}_2(p)^{-1} &\bm{0}&\bm{0}\\
	\bm{0}& \bm{0}& \frac{1}{\lambda} \mathbf{I}_f(\bm{\theta}_2)^{-1}  &\bm{0}   \\
	\bm{0}& \bm{0}& \bm{0}&  \frac{1}{cp}\mathbf{I}_g(\bm{\theta}_1)^{-1}
\end{array}
\right)\right).
\]
	as $K_2 \rightarrow +\infty$. $\mathbf{I}_1(\lambda)=\frac{1}{\lambda}$ is the information matrix for poisson distribution. $\mathbf{I}_2(p)=\frac{1}{p(1-p)}$ is the information matrix for bernoulli distribution. $\mathbf{I}_f(\theta_2)$ is the information matrix for distribution $F_{\theta_2}$. $\mathbf{I}_g(\theta_1)$ is the information matrix for distribution $G_{\theta_1}$. c is the limit of $\frac{T}{K_2}$ as $K_2\rightarrow +\infty$.

\end{theorem}

The proof of the theorem is given in Supporting Information Web Appendix A. This theorem provides us with the asymptotic normality of the MLE estimator. Our next step is to derive an estimator for AUC and the associated asymptotic property in section~\ref{sec:AFRCO} below.

\subsection{The Confidence Interval for the AUC of the AFROC curve}
\label{sec:AFRCO}

It has been shown that the AUC of the AFROC curve has the following expression, 
\[
	AUC=p(1-e^{-\lambda})P(Y_{is}>\max \limits_{1 \leq k \leq m_j}{X_{jk}}|m_j>0,L_{is}=1)+\frac{(1+p)e^{-\lambda}}{2}.
\]
The following theorem gives a simpler formula for AUC:

\begin{theorem}
	\label{constructionCIAUC}
	Under Assumption $\bm{A}_1$ to $\bm{A}_5$, the AUC of the AFROC curve can be expressed as follows:
	\[
	AUC=g(p,\lambda,\bm{\theta}_1,\bm{\theta}_2)\]
	\[=p(1-e^{-\lambda})P(Y_{is}>\max \limits_{1\leq k\leq m_j}{X_{jk}}|m_j>0,L_{is}=1)+\frac{(1+p)e^{-\lambda}}{2}\]
	\[=pe^{-\lambda}E_{Y^*}(e^{\lambda F_{\bm{\theta}_2}(Y^*)}-1)+\frac{(1+p)e^{-\lambda}}{2},
	\]
	where $Y^*\sim G_{\bm{\theta}_1}$, $G_{\bm{\theta}_1}$ is the cdf of the confidence scores of the true positive locations.

\end{theorem}

The proof is shown in Supporting Information Web Appendix B. The estimator of AUC, $\widehat{AUC}$, can be derived by plugging in the MLE estimator $(\hat{p},\hat{\lambda},\hat{\bm{\theta}}_1,\hat{\bm{\theta}}_2)$ into $g(p,\lambda,\bm{\theta}_1,\bm{\theta}_2)$. Combined with Theorem~\ref{asmptoticprop}, the asymptotic distribution of $\widehat{AUC}$ can be derived by delta method. Denote $\bm{\gamma}=(p,\lambda,\bm{\theta}_1,\bm{\theta}_2)$ and $\hat{\bm{\gamma}}=(\hat{p},\hat{\lambda},\hat{\bm{\theta}}_1,\hat{\bm{\theta}}_2)$. Let $\Sigma(\bm{\gamma})$ be the covariance matrix for asymptotic distribution of $\bm{\gamma}$. The asymptotic variance of $\widehat{AUC}$ can be estimated by $\widehat{\sigma_{AUC}}^2=\frac{\partial g(\hat{\bm{\gamma}})}{\partial \bm{\gamma}}\Sigma(\hat{\bm{\gamma}})\frac{\partial g(\hat{\bm{\gamma}})}{\partial \bm{\gamma}}^T$. The confidence interval for AUC of level $1-\alpha$ can be constructed as \[\left(\widehat{AUC}-\frac{\widehat{\sigma_{AUC}}\Phi^{-1}(1-\frac{\alpha}{2})}{\sqrt{N}},\widehat{AUC}+\frac{\widehat{\sigma_{AUC}}\Phi^{-1}(1-\frac{\alpha}{2})}{\sqrt{N}}\right),\]
  where $\Phi(\cdot)$ is the cumulative distribution function of standard normal distribution.

\subsection{The Confidence Interval for the lesion localization fraction at a fixed false positive fraction}


On an AFROC curve, the ordinate, the lesion localization fraction (LLF), represents the lesion level true positive rate, and the abscissa, the false positive fraction (FPF), represents the subject level false positive rate. The following theorem shows the relationship between the LLF at a given FPF and the parameters $(p,\lambda,\bm{\theta}_1,\bm{\theta}_2)$:

\begin{theorem}
	\label{LLFgivenFPF}
	Under Assumption $\bm{A}_1$ to $\bm{A}_5$,
	for a given false positive fraction (FPF) q, the lesion location fraction $LLF_q$ is given by:
	
	\begin{equation}
		LLF_q(p,\lambda,\bm{\theta}_1,\bm{\theta}_2)=p(1-G_{\bm{\theta}_1}(F_{\bm{\theta}_2}^{-1}(1+\frac{log(1-q)}{\lambda}))),
	\end{equation}
where $G_{\bm{\theta}_1}$ is the cdf of the confidence scores of the true positive locations and $F_{\bm{\theta}_2}$ is the cdf of the confidence scores of the false positive locations on negative subjects. 
\end{theorem}

The proof is given in Supporting Information Web Appendix C. By plugging in the MLE estimator $(\hat{p},\hat{\lambda},\hat{\bm{\theta}}_1,\hat{\bm{\theta}}_2)$ into $LLF_q(p,\lambda,\bm{\theta}_1,\bm{\theta}_2)$, we derive the estimator $\widehat{LLF_q}$. Denote $\bm{\gamma}=(p,\lambda,\bm{\theta}_1,\bm{\theta}_2)$ and $\hat{\bm{\gamma}}=(\hat{p},\hat{\lambda},\hat{\bm{\theta}}_1,\hat{\bm{\theta}}_2)$. Let $\Sigma(\bm{\gamma})$ be the covariance matrix for asymptotic distribution of $\bm{\gamma}$. The asymptotic variance of $\widehat{LLF_q}$ can be estimated by $\widehat{\sigma_{LLF_q}}^2=\frac{\partial LLF_q(\hat{\bm{\gamma}})}{\partial \bm{\gamma}}\Sigma(\hat{\bm{\gamma}})\frac{\partial LLF_q(\hat{\bm{\gamma}})}{\partial \bm{\gamma}}^T$. The confidence interval for $LLF_q$ of level $1-\alpha$ can be constructed as \[\left(\widehat{LLF_q}-\frac{\widehat{\sigma_{LLF_q}}\Phi^{-1}(1-\frac{\alpha}{2})}{\sqrt{N}},\widehat{LLF_q}+\frac{\widehat{\sigma_{LLF_q}}\Phi^{-1}(1-\frac{\alpha}{2})}{\sqrt{N}}\right),\]
	where $\Phi(\cdot)$ is the cumulative distribution function of standard normal distribution.

\begin{figure}
	\centerline{\includegraphics[width=2.25in]{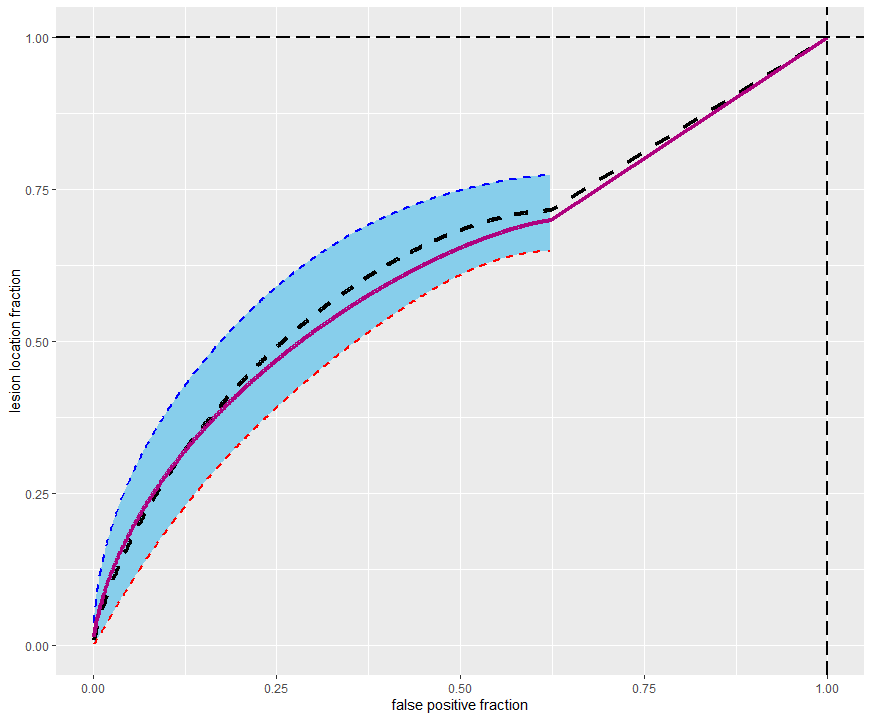}}
	\caption{An example of the pointwise confidence interval for an AFROC curve}
	\label{f:CIbandexample}
\end{figure}

By constructing the confidence interval of $LLF_q$ for each q, we can construct a point-wise confidence interval for the AFROC curve. When q is near to zero, $LLF_q$ is small and the lower bound of $LLF_q$ may be lower than 0. We can first take a logit transformation of $LLF_q$ to guarantee that the lower bound is positive. An example of the pointwise confidence interval is shown in Figure \ref{f:CIbandexample}. The purple solid line is the true AFROC curve while the black dotted line is the estimated AFROC curve. The upper bound and lower bound of the pointwise confidence interval for the AFROC curve are the dotted blue line and the dotted red line, respectively. The blue area between the red dotted line and the blue dotted line is the pointwise confidence interval. The reason why we only construct the confidence interval from 0 to $1-\hat{P}(m_j=0)$ is that the second part of the AFROC curve (the straight line in Figure~\ref{f:AFROCexample}) has no clinical meaning. The straight line in the AFROC curve is used to form the second component of AUC, $S_2$, while the points on this straight line are not meaningful.

\subsection{The Confidence Sets for Multiple Indices}

In a clinical study, there may be multiple indices that are thought to be important. Different indices can capture different aspects of the performance of the diagnostic methods. For example, in a ROC study, \citet{yin2021joint} considers both the ROC-AUC and the Youden index. In our example in section \ref{s:application}, both the AUC of the AFROC curve and the lesion discovery rate (defined as $\text{P}(L_{is})=1$) are considered important. Our method can be directly used for the construction of confidence sets for multiple indices.

Suppose we consider M indices $h_1(p,\lambda_1,\lambda_2,\bm{\theta}_1,\bm{\theta}_2,\bm{\theta}_3),\dots,h_M(p,\lambda_1,\lambda_2,\bm{\theta}_1,\bm{\theta}_2,\bm{\theta}_3)$. Denote $\bm{\gamma}=(p,\lambda_1,\lambda_2,\bm{\theta}_1,\bm{\theta}_2,\bm{\theta}_3)$. The estimator, $(\hat{h}_1,\hat{h}_2,\dots,\hat{h}_M)$, can be derived by plugging in the MLE estimator $(\hat{p},\hat{\lambda},\hat{\bm{\theta}}_1,\hat{\bm{\theta}}_2)$ into  $(h_1,h_2,\dots,h_M)$.
By the delta method, the asymptotic distribution of $\hat{\vec{h}}=(\hat{h}_1,\dots,\hat{h}_M)$ follows a multivariate normal distribution,
\[
	\sqrt{K_2}\left( 
	\begin{array}{c}   
		\widehat{h}_1-h_1 \\  
		\dots \\  
		\widehat{h}_M-h_M
	\end{array}
	\right)\xrightarrow{d}\mathbf{N}\left(\bm{0},
	\frac{\partial \vec{h}(\bm{\gamma})}{\partial \bm{\gamma}} \Sigma(\bm{\gamma}){\frac{\partial \vec{h}(\bm{\gamma})}{\partial \bm{\gamma}}}^T
	\right)
\]
where $\Sigma(\bm{\gamma})$ is the covariance matrix for the asymptotic distribution of $\hat{\bm{\gamma}}$. Then the confidence sets for $\vec{h}$ can be constructed by:
\[
	K_2(\hat{\vec{h}}-\vec{h})^T\{
	\frac{\partial\vec{h}(\hat{\gamma})}{\partial \bm{\gamma}} \Sigma(\hat{\bm{\gamma}}){\frac{\partial\vec{h}(\hat{\gamma})}{\partial \bm{\gamma}}}^T\}^{-1}(\hat{\vec{h}}-\vec{h})\leq \chi^{-1}_{M-1}(1-\alpha),
\]
where $\chi^{-1}_{M-1}(1-\alpha) $ is the $1-\alpha$ quantile of the chi-square distribution with M-1 degree of freedom and the covariance matrix is estimated by plugging in the consistent MLE estimator.

\section{Simulation}
\label{s:simulation}

In this section, we carry out two simulations to evaluate the finite sample performance of our method. The first simulation is used to assess the coverage rate and the average length of the confidence interval for the AUC of the AFROC curve. The second simulation is used to assess the coverage rate and the average length of the confidence interval for the lesion location fraction (LLF) at a fixed false positive fraction (FPF).

In our simulation study, we generate the FROC datasets allowing data to be correlated within the same subject. All parameters in the simulation are set to be in the common range of clinical practice. The number of positive subjects and the number of negative subjects are set to be $n$ and $m$, respectively. We choose three settings including $n=m=50$, $n=m=100$ and $n=m=200$, which are the common sample sizes in clinical practice. The number of lesions on each positive subject is set to be $t=2$. The probability for each lesion to be found, i.e. $P(L_{is}=1)$ is set to be $p_0$, where we take $p_0$ to be 0.6 or 0.8. The number of false positive marks on a normal subject is set to follow a poisson distribution with parameter $\lambda$, where we set $\lambda$ to be 0.5, 1, or 1.5. The confidence score is set to follow a normal distribution. To allow correlation within the same subject, we consider a random effect model. A random effect $\sigma^{(1)}_i\sim \text{N}(0,\sigma_{01}^2)$ is independently generated for positive subject i. Conditioning on the random effect $\sigma^{(1)}_i$, the confidence scores of the true positive locations on subject i follow normal distributions with mean $\mu_i^{(1)}=\mu_1+\sigma^{(1)}_i$ and standard deviation $\sigma_1$. A similar random effect $\sigma^{(2)}_j\sim \text{N}(0,\sigma_{02}^2)$ is also independently generated for negative subject j. Conditioning on the random effect $\sigma^{(2)}_j$, the confidence scores of the false positive locations on negative subject j follow normal distributions with mean $\mu_j^{(2)}=\mu_2+\sigma^{(2)}_j$ and standard deviation $\sigma_2$. In this study, $\mu_1$ is set to be 2, $\mu_2$ is set to be 1 and $\sigma_1=\sigma_2=1$. We choose different $\sigma_{01}$ and $\sigma_{02}$ to test the effect of the independence assumption on the coverage rate of the confidence interval. We choose $\sigma_{01}=\sigma_{02}=0$ which means no correlation within the same subject and $\sigma_{01}=\sigma_{02}=0.3$ for mild correlation within the same subject. For each setting, we generate 10,000 FROC datasets.

\subsection{Simulation One}
 In simulation one, we derive the point estimates for the AUC of the AFROC curve using our method and the empirical method respectively. The confidence interval for AFROC-AUC is constructed using our main results or the bootstrap method. The coverage rate and the average length of the confidence intervals are reported and compared. The simulation results are shown in Table \ref{t:simulation1} and Table \ref{t:simulation2}.

From the simulation results, we conclude that our method is valid even under mild correlation within the same subject. As can be seen from the simulation results, the CI coverage rates are similar for the two methods when data is uncorrelated within the same subject. However, the average CI length of our method is generally shorter than the empirical method's. When the independence assumption is not true, the CI coverage rates are not greatly affected. This is reasonable since the within subject correlation in our simulation is mild, which we believe is consistent with the clinical practice of the computer observers.


\begin{table}
	\caption{Simulation result one for the proposed method. Coverage represents the coverage rate of the estimated 95\% confidence intervals. Length represents average length of the confidence intervals.}
	\label{t:simulation1}
	\begin{center}
			\begin{tabular}{lllcccccc}
			\Hline
			&      &     & \multicolumn{2}{c}{m=n=50} & \multicolumn{2}{c}{m=n=100} &\multicolumn{2}{c}{m=n=200} \\
\hline
			$\lambda$& $p_0$&  $\sigma_{01}$ &  coverage & length &    coverage  &  length & coverage &  length \\
		\hline
			0.5 & 0.8 & 0   & 0.9402 & 0.1425 & 0.9430 & 0.1012 & 0.9477 & 0.0718 \\
			&     & 0.3 & 0.9439 & 0.1439 & 0.9487 & 0.1023 & 0.9495 & 0.0724 \\
			& 0.6 & 0   & 0.9462 & 0.1553 & 0.9451 & 0.1103 & 0.9452 & 0.0781 \\
			&     & 0.3 & 0.9451 & 0.1560 & 0.9507 & 0.1107 & 0.9483 & 0.0784 \\
		\hline
			1   & 0.8 & 0   & 0.9418 & 0.1628 & 0.9471 & 0.1155 & 0.9491 & 0.0818 \\
			&     & 0.3 & 0.9404 & 0.1638 & 0.9443 & 0.1162 & 0.9447 & 0.0823 \\
			& 0.6 & 0   & 0.9426 & 0.1686 & 0.9485 & 0.1195 & 0.9483 & 0.0846 \\
			&     & 0.3 & 0.9438 & 0.1689 & 0.9444 & 0.1197 & 0.9451 & 0.0848 \\
\hline
			1.5 & 0.8 & 0   & 0.9454 & 0.1701 & 0.9465 & 0.1206 & 0.9457 & 0.0854 \\
			&     & 0.3 & 0.9439 & 0.1707 & 0.9419 & 0.1210 & 0.9432 & 0.0857 \\
			& 0.6 & 0   & 0.9459 & 0.1703 & 0.9493 & 0.1207 & 0.9479 & 0.0854 \\
			&     & 0.3 & 0.9459 & 0.1702 & 0.9461 & 0.1206 & 0.9471 & 0.0854\\
			\hline
		\end{tabular}
	\end{center}
\end{table}

\begin{table}
	\caption{Simulation result one for the empirical method. Coverage represents the coverage rate of the estimated 95\% confidence intervals. Length represents the average length of the confidence intervals.}
	\label{t:simulation2}
	\begin{center}
		\begin{tabular}{lllcccccc}
			\Hline
			&      &     & \multicolumn{2}{c}{m=n=50} & \multicolumn{2}{c}{m=n=100} &\multicolumn{2}{c}{m=n=200} \\
			\hline
			$\lambda$& $p_0$&  $\sigma_{01}$ &  coverage & length &    coverage  &  length & coverage &  length \\
			\hline
		0.5 & 0.8 & 0   & 0.9438 & 0.1474 & 0.9435 & 0.1042 & 0.9472 & 0.0737 \\
		&     & 0.3 & 0.9403 & 0.1486 & 0.9450 & 0.1051 & 0.9461 & 0.0745 \\
		& 0.6 & 0   & 0.9416 & 0.1603 & 0.9469 & 0.1137 & 0.9488 & 0.0805 \\
		&     & 0.3 & 0.9435 & 0.1613 & 0.9440 & 0.1144 & 0.9472 & 0.0809 \\
		\hline
		1   & 0.8 & 0   & 0.9426 & 0.1712 & 0.9468 & 0.1213 & 0.9499 & 0.0858 \\
		&     & 0.3 & 0.9436 & 0.1729 & 0.9457 & 0.1224 & 0.9454 & 0.0866 \\
		& 0.6 & 0   & 0.9466 & 0.1780 & 0.9468 & 0.1261 & 0.9483 & 0.0892 \\
		&     & 0.3 & 0.9465 & 0.1788 & 0.9510 & 0.1266 & 0.9487 & 0.0896 \\
		\hline
		1.5 & 0.8 & 0   & 0.9436 & 0.1811 & 0.9495 & 0.1281 & 0.9481 & 0.0907 \\
		&     & 0.3 & 0.9439 & 0.1827 & 0.9503 & 0.1294 & 0.9470 & 0.0916 \\
		& 0.6 & 0   & 0.9484 & 0.1814 & 0.9479 & 0.1286 & 0.9493 & 0.0910 \\
		&     & 0.3 & 0.9464 & 0.1822 & 0.9490 & 0.1292 & 0.9468 & 0.0914\\
			\hline
		\end{tabular}
	\end{center}
\end{table}

\subsection{Simulation Two}
In simulation two, we estimate the lesion location fraction (LLF) at a fixed false positive fraction (FPF). The data generating process is the same as that described above. We choose to estimate LLF given $\text{FPF}=q$ and q is set to be 0.1. Since there are no existing methods for the inference of the LLF at a fixed FPF, here we only present the coverage rate and average length of the confidence interval by our method in Supporting Information Web Table 1. As can be seen from the results, the performance for estimating the LLF at a fixed FPF is similar to that of the AUC. The speed of convergence is slower compared to that of the AUC, but the coverage rate still achieves about 94\% when the sample size is 200. The effect of the within subject correlation is also similar to that of the AUC. The coverage rate is lower under mild within subject correlation, but still achieves about 92\% in most settings. The effect of violation of the independence assumption will be larger when $\lambda$ and p are large. This is because when $\lambda$ and p are small, most subjects contain at most one true positive or false positive location, which means that the correlation within the same subject does not have a great effect on the data. We conclude that our method is also valid for estimating LLF at a fixed FPF.

In conclusion, our method is valid for constructing confidence intervals for AUC and the LLF at a given FPF. It outperforms the empirical method when the model assumption is correct. If the independence assumption is wrong, it is not greatly affected under mild within subject correlation. An important advantage of our method is that it is a unified method which can simultaneously estimate the smooth curve, derive the confidence interval for summary indices, and derive the confidence interval for the true positive rate at a fixed false positive rate while existing methods can only do one of these tasks. As a result, we conclude that our method is valid and has its own advantages compared to the existing methods.

\section{Application}
\label{s:application}

The FROC data in this real case study comes from an unpublished clinical trial of a Software as a Medical Device (SaMD) by InferRead CT Lung, Infervision, Beijing, China. The effectiveness of this SaMD has been approved by National Medical Products Administration (NMPA). The SaMD is designed to aid the readers in finding pulmonary lesions but can also work alone. The primary goal of this study is to evaluate and compare the accuracy of readers with and without the SaMD for pulmonary lesions diagnosis but the accuracy of the SaMD working alone is also concerned. This study consists of a multi-reader-multi-case reader study and a SaMD working alone study. In our application, we use the data of SaMD working alone to evaluate the SaMD standalone performance. 

The data used in this section consists of 344 subjects. The gold standard is set by three experienced experts. The first and the second expert independently read the image and record the locations, sizes, and categories of the lesions. The last expert acts as an arbitrator. If the first two experts have different opinions, the last expert makes the final decision. In the SaMD working alone study, the SaMD reports the locations, sizes, and categories of the detected lesions. In this application, the subgroup of lesions over 5mm in size is analyzed. Subjects with any category of lesions larger than 5mm are classified as positive subjects, which results in 120 positive subjects and 224 negative subjects. There are totally 201 lesions on the positive subjects by the gold standard which means there are on average about 1.7 lesions on each positive subject. The SaMD standalone data records the suspicious locations with a diameter larger than 5mm and their corresponding confidence scores. The confidence scores of the SaMD are positive continuous variables ranging from 0.75 to 1. The distribution of the confidence scores is difficult to model directly. However, it can be shown in Supporting Information Web Appendix D that a monotonous transform of the TP confidence scores and the FP confidence scores will not change the AUC of the AFROC curve and the LLF at a fixed FPF, so we first rescale the confidence scores to a range of 0 to 1. Beta distributions are used to model the rescaled confidence scores of true positive locations and false positive locations. Kolmogorov-Smirnov tests are used to test whether the fitted beta distributions are consistent with the data. Kolmogorov-Smirnov tests accept null hypotheses with p values of 0.1171 and 0.912. The MLE estimates for beta distribution parameters are $\alpha_1=2.575$, $\beta_1=0.627$ for TP scores and $\alpha_2=1.234$, $\beta_2=1.560$ for FP scores. Histograms for confidence scores of TP locations and FP locations (on normal subjects) are shown in Supporting Information Web Figure 1 and 2. The fitted distributions are shown in Supporting Information Web Figure 3 and 4. 

\begin{figure}
	\centerline{\includegraphics[width=2.25in]{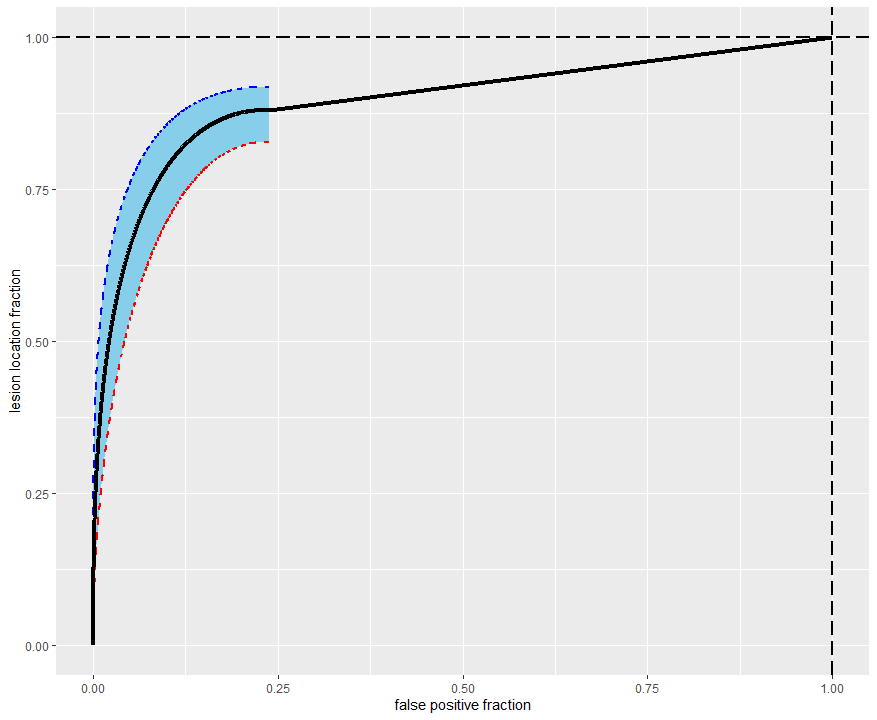}}
	\caption{Estimated AFROC curve and its pointwise confidence interval.}
	\label{f:applicationafroc}
\end{figure}

The point estimate and 95\% confidence interval for the AUC of the AFROC curve given by our method are 0.8955 (0.8649,0.9262). The point estimate and 95\% confidence interval for the AUC of the AFROC curve given by the empirical method (variance calculated by the bootstrap method) are 0.8996 (0.8687,0.9304). The smooth AFROC curve and its pointwise confidence interval given by our method are shown in Figure~\ref{f:applicationafroc}. 

\begin{figure}
	\begin{center}
		\centerline{\includegraphics[width=3.75in]{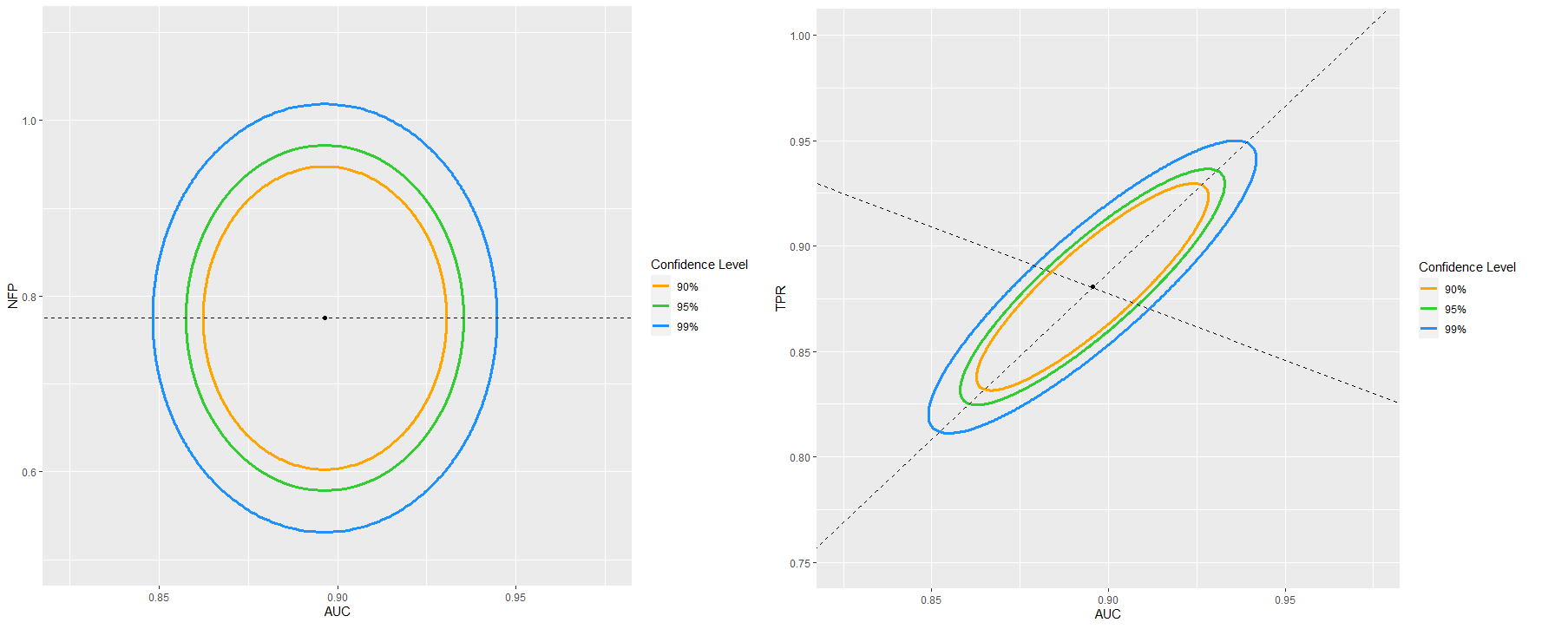}}
	\end{center}
	\caption{Confidence sets for AUC and $\lambda_2$ (left) and confidence sets for AUC and p (right). }
	\label{f:CIarea}
\end{figure}

In this particular example, we find that a single index such as the AUC of the AFROC curve may not be enough to characterize the performance of the SaMD. In this example, the average number of FP is 0.775 on positive subjects and 0.272 on negative subjects. The average number of FP on the positive subjects is much larger than that on the negative subjects. This is reasonable since positive subjects and negative subjects may have different features so the average number of the false positive locations may be different. AFROC-AUC does not consider the information of FP locations on positive subjects. Other indices like the AUC of the FROC curve mix FP locations on the negative subjects and the positive subjects together so it can not characterize the special feature of this FROC data. This example shows that the AUC and the average number of FP locations on the positive subjects actually characterize different aspects of the performance of the SaMD so it is necessary to evaluate multiple indices at the same time. Our method can be directly used for constructing the confidence sets for multiple indices, which has not been studied before. An example of the confidence sets for $(\text{AUC},\lambda_2)$ in our example is shown in Figure~\ref{f:CIarea}, where $\lambda_2$ is the average number of FP marks on the positive subjects. 

In this study, the TP discovery rate, $p$ is also considered important since the TP discovery rate is the largest lesion location fraction AI can achieve no matter how the threshold changes. This SaMD is designed to aid the human readers, not to replace the human readers. As a result, the detection of lesion locations in the first stage, which is represented by p, is of great importance. We also shows the confidence sets for $(\text{AUC},p)$ in Figure ~\ref{f:CIarea}.

Note that $(\text{AUC},\lambda_2)$ and $(\text{AUC},p)$ are not the indices concerned in every study. Which index is chosen should depend on the design of the diagnostic tools and the goal of the study. However, this example shows the advantage of our method in constructing confidence sets for multiple indices. If there are other indices concerned, our method can also be applied directly.

\section{Discussion}
\label{s:discuss}

In this article, we propose a unified method based on the IDCA model for the inference of FROC data. An MLE estimator is proposed. The unique feature of FROC data brings a problem of random sample size, which means that the number of scores given by the reader is informative. By adjusting the regular proof of the normality for the MLE estimator, we derive the asymptotic normality of our MLE estimator. Using this estimator, we can simultaneously estimate a smooth FROC-type curve, construct the confidence interval for LLF at a given FPF, and construct confidence intervals for summary indices like the AUC of the AFROC curve. We note that though only the formula for AFROC is derived in section~\ref{s:result}, our method can be also be used for general FROC-type curves and summary indices. As the generalization of our method to other curves and summary indices is straight forward, we do not show the results for all of them in detail. The AFROC curve and its AUC are chosen as the example in section~\ref{s:result} because they are recommended in \citet{chakraborty2017observer} and have been widely used in diagnostic research by U.S. Food and Drug Administration (\citet{fda1} and \citet{fda2}, for example). Our method has several advantages. First, it has a valid theoretical guarantee compared to the existing methods. Second, it is the first method proposed for the inference of lesion location fraction (LLF) at a fixed false positive (FPF) fraction as far as we know. Third, our method can also be used for the estimation and the inference of multiple indices. In conclusion, our method can serve as a useful tool for the analysis of FROC data, especially in the field of Software as a Medical Device (SaMD).

There are still some limitations to this work. The first limitation is that the independence assumption $\bm{A}_2$, $\bm{A}_3$, $\bm{A}_4$ and $\bm{A}_5$ may be too strong in some situation and is difficult to justify \citep{bandos2009area}. However, we make the same argument as the one in \citet{edwards2002maximum} that independence assumption may be more credible for computerized observers such as the SaMD, which is the same case for our application. More importantly, our simulation results show that even if data within the same subject are correlated, our method can still achieve good performance. An explanation for this simulation result is that the average number of the FP locations recorded by the reader on each subject is generally not large in clinical practice so the performance of our method is not greatly affected by the within subject correlation. Another limitation is that we only discuss about continuous data in this article, which is common for the SaMD standalone studies. However, ordinal data is also common for human reader studies. It is simple to extend our method to ordinal data for point estimation by using latent model (this is in fact the work of \citet{edwards2002maximum}), but the construction of confidence interval may need more theoretical studies.

For our future work, we want to improve our method to address the above limitations. First, we plan to relax the independence assumption by using a random effect model. The theoretical study is undergoing. Second, we want to extend our method to ordinal data and study the regression methods for the FROC-type curves by adding subject-related or reader-related covariates. Finally, we plan to study the asymptotic properties of the nonparametric estimates of the FROC-type curves and related summary indices.


\backmatter


\section*{Acknowledgements}


This research is funded by National Natural Science Foundation of China (82173623). We would like to thank the Beijing Infervision Technology Co. Ltd. for providing the dataset of the SaMD example.\vspace*{-8pt}


%

\bibliographystyle{biom} 
\bibliography{mybib}

@article{edwards2002maximum,
  title={Maximum likelihood fitting of FROC curves under an initial-detection-and-candidate-analysis model},
  author={Edwards, Darrin C and Kupinski, Matthew A and Metz, Charles E and Nishikawa, Robert M},
  journal={Medical physics},
  volume={29},
  number={12},
  pages={2861--2870},
  year={2002},
  publisher={Wiley Online Library}
}

@article{chakraborty1989maximum,
  title={Maximum likelihood analysis of free-response receiver operating characteristic (FROC) data},
  author={Chakraborty, Dev P},
  journal={Medical physics},
  volume={16},
  number={4},
  pages={561--568},
  year={1989},
  publisher={Wiley Online Library}
}

@book{chakraborty2017observer,
  title={Observer performance methods for diagnostic imaging: foundations, modeling, and applications with r-based examples},
  author={Chakraborty, Dev P},
  year={2017},
  publisher={CRC Press}
}

@inproceedings{bunch1977free,
  title={A free response approach to the measurement and characterization of radiographic observer performance},
  author={Bunch, Philip C and Hamilton, John F and Sanderson, Gary K and Simmons, Arthur H},
  booktitle={Application of Optical Instrumentation in Medicine VI},
  volume={127},
  pages={124--135},
  year={1977},
  organization={International Society for Optics and Photonics}
}

@article{chakraborty1990free,
  title={Free-response methodology: alternate analysis and a new observer-performance experiment.},
  author={Chakraborty, Dev P and Winter, LH},
  journal={Radiology},
  volume={174},
  number={3},
  pages={873--881},
  year={1990}
}

@article{bandos2009area,
  title={Area under the free-response ROC curve (FROC) and a related summary index},
  author={Bandos, Andriy I and Rockette, Howard E and Song, Tao and Gur, David},
  journal={Biometrics},
  volume={65},
  number={1},
  pages={247--256},
  year={2009},
  publisher={Wiley Online Library}
}

@article{chakraborty2004observer,
  title={Observer studies involving detection and localization: modeling, analysis, and validation},
  author={Chakraborty, Dev P and Berbaum, Kevin S},
  journal={Medical physics},
  volume={31},
  number={8},
  pages={2313--2330},
  year={2004},
  publisher={Wiley Online Library}
}

@inproceedings{samuelson2006comparing,
  title={Comparing image detection algorithms using resampling},
  author={Samuelson, Frank W and Petrick, Nicholas},
  booktitle={3rd IEEE International Symposium on Biomedical Imaging: Nano to Macro, 2006.},
  pages={1312--1315},
  year={2006},
  organization={IEEE}
}

@misc{
fda1,
author = {FDA},
title = {PMA P120004: FDA Summary of Safety and Effectiveness Data[EB/OL]},
howpublished = {Website},
year = {2019},
note = {\url{https://www.accessdata.fda.gov/cdrh\_docs/pdf12/p120004b.pdf}}
}

@misc{
fda2,
author = {FDA},
title = {FDA -P130020- Summary of Safety and Effectiveness Data(SSED)[EB/OL]},
howpublished = {Website},
year = {2019},
note = {\url{https://www.accessdata.fda.gov/cdrh\_docs/pdf13/P130020b.pdf}}
}

@article{popescu2011nonparametric,
  title={Nonparametric signal detectability evaluation using an exponential transformation of the FROC curve},
  author={Popescu, Lucretiu M},
  journal={Medical physics},
  volume={38},
  number={10},
  pages={5690--5702},
  year={2011},
  publisher={Wiley Online Library}
}

@article{chakraborty2008validation,
  title={Validation and statistical power comparison of methods for analyzing free-response observer performance studies},
  author={Chakraborty, Dev P},
  journal={Academic radiology},
  volume={15},
  number={12},
  pages={1554--1566},
  year={2008},
  publisher={Elsevier}
}

@article{chakraborty2006search,
  title={A search model and figure of merit for observer data acquired according to the free-response paradigm},
  author={Chakraborty, Dev P},
  journal={Physics in Medicine \& Biology},
  volume={51},
  number={14},
  pages={3449},
  year={2006},
  publisher={IOP Publishing}
}

@article{winawer2007colorectal,
  title={Colorectal cancer screening},
  author={Winawer, Sidney J},
  journal={Best practice \& research Clinical gastroenterology},
  volume={21},
  number={6},
  pages={1031--1048},
  year={2007},
  publisher={Elsevier}
}

@article{black2000anatomic,
  title={Anatomic extent of disease: a critical variable in reports of diagnostic accuracy},
  author={Black, William C},
  journal={Radiology},
  volume={217},
  number={2},
  pages={319--320},
  year={2000},
  publisher={Radiological Society of North America}
}

@book{zhou2009statistical,
  title={Statistical methods in diagnostic medicine, Second Edition},
  author={Zhou, Xiao-Hua and McClish, Donna K and Obuchowski, Nancy A},
  year={2011},
  publisher={John Wiley \& Sons}
}

@article{nagel1998analysis,
  title={Analysis of methods for reducing false positives in the automated detection of clustered microcalcifications in mammograms},
  author={Nagel, Rufus H and Nishikawa, Robert M and Papaioannou, John and Doi, Kunio},
  journal={Medical Physics},
  volume={25},
  number={8},
  pages={1502--1506},
  year={1998},
  publisher={Wiley Online Library}
}

@article{yin2021joint,
  title={Joint inference about the AUC and Youden index for paired biomarkers},
  author={Yin, Jingjing and Samawi, Hani and Tian, Lili},
  journal={Statistics in Medicine},
  year={2021},
  publisher={Wiley Online Library}
}

@article{miller1969froc,
  title={The FROC curve: A representation of the observer's performance for the method of free response},
  author={Miller, Harold},
  journal={The Journal of the Acoustical Society of America},
  volume={46},
  number={6B},
  pages={1473--1476},
  year={1969},
  publisher={Acoustical Society of America}
}


\section*{Supporting Information}

Web Appendices, Tables, and Figures referenced in Sections 3-5 are available with this paper at the Biometrics website on Wiley Online Library.\vspace*{-8pt}


\label{lastpage}

\end{document}